\documentclass[aps,pre,showpacs,preprint]{revtex4}

\usepackage{amsmath}
\usepackage{graphicx}
\usepackage{color}

\begin{document}

\title{Influence of spatially modified tissue on atrial fibrillation
  patterns: Insight from solutions of the FitzHugh-Nagumo equations}

\author{Claudia Lenk$^{1}$}
\email{claudia.lenk@tu-ilmenau.de}
\author{Mario Einax$^{1}$}
\author{Philipp Maass$^{2}$}
\email{pmaass@uos.de}
\homepage{http://www.statphys.uni-osnabruck.de}
\affiliation{ $^{1}$Institut f\"ur Physik, Technische
Universit\"at Ilmenau, 98684 Ilmenau, Germany\\
$^{2}$Fachbereich Physik, Universit\"at Osnabr\"uck,
Barbarastra{\ss}e 7, 49069 Osnabr\"uck, Germany }

\date{6 April 2010} 

\begin{abstract} We study the interplay between traveling action
  potentials and spatial inhomogeneities in the FitzHugh-Nagumo model
  to investigate possible mechanisms for the occurrence of
  fibrillatory states in the atria of the heart. Different dynamical
  patterns such as ectopic foci, localized and meandering spiral waves
  are found depending on the characteristics of the inhomogeneities.
  Their appearance in dependence of the size and strength of the
  inhomogeneities is quantified by phase diagrams. Furthermore it is
  shown that regularly paced waves in a region R, that is connected by
  a small bridge connection to another region L with perturbing waves
  emanating from an additional pacemaker, can be strongly disturbed,
  so that a fibrillatory state emerges in region R after a transient
  time interval. This finding supports conjectures that fibrillatory
  states in the right atrium can be induced by self-excitatory
  pacemakers in the left atrium.
\end{abstract}


\maketitle

\section{Introduction} Atrial fibrillation (AF) is the most frequently
appearing heart arrhythmia since it accounts for one third of all
hospitalizations caused by heart arrhythmia in the industrialized
countries \cite{Europace}. During AF the electric conduction system of
the heart is disturbed and an increased rate of activation by a factor
of 3-12 compared to normal sinus rhythm occurs. Special
spatio-temporal patterns of the electric potential like spiral waves,
mother waves or ectopic foci are thought to be underlying generating
mechanisms of AF \cite{Nattel,Nattel2,Nattel3,Mandapati}. These
patterns are often located near physiologically modified regions of
the heart tissue in the left atrium
\cite{Wu,Mandapati,Sahadevan,Sanders}. The question hence arises, how
these physiologically modified regions can be responsible for the
generation of spiral waves or ectopic foci and how they influence the
properties of these patterns.

To tackle these questions, we study generating mechanism for AF on the
basis of the FitzHugh-Nagumo model \cite{FHN}, which is a simple model
for action potential generation and propagation. By modeling
physiologically modified regions using a spatial variation of the
parameters characterizing cell properties like excitability or resting
state stability, we calculate phase diagrams, which specify the type
of spatio-temporal excitation pattern in dependence of the extent of
the modified region and the strength of the modification. Thereupon we
investigate how self-excitatory sources as spiral waves or ectopic
foci with rather regular dynamics in one region can induce irregular,
fibrillatory excitation patterns in some other region. Irregular,
fibrillatory states are often observed in the right atrium
\cite{Sahadevan, Sanders, Lazar} and it was conjectured that these are
caused by the perturbation of regular waves generated by the sinus
node by waves emanating from an additional pacemaker like a spiral
wave or ectopic foci in the left atrium.

\section{Model}
\label{ch:model} 

The FitzHugh-Nagumo (FHN) equations \cite{FHN} are a set of two
coupled nonlinear ordinary differential equations, which describe
excitable media via an inhibitor-activator mechanism. They were
originally developed by searching for a simplified version of the
Hodgkin-Huxley equations for electric pulse propagation along nerves
\cite{HH}. When combined with a spatial diffusion term, the equations
are
\begin{eqnarray}
  \frac{\partial u}{\partial t} &=& D\left(\frac{\partial^2
      u}{\partial x^2}+\frac{\partial^2 u}{\partial y^2}\right)
      + c(v+u-\frac{u^3}{3}+z) \nonumber \, \\
  \frac{\partial v}{\partial t} &=& -\frac{1}{c}(u-a+b v)\;.
\label{eq:FHN}
\end{eqnarray}
This set of partial differential equations serves as a prototype for a
large variety of reaction-diffusion systems, which occur, for example,
in chemical reactions as the Bhelousov-Zhabotinsky reaction
\cite{Tobias,Baer} or the catalysis of carbon monoxide
\cite{Jakubith,Ertl}, in population dynamics \cite{Clerc}, in biology
in connection with aggregation processes \cite{Lee} or plancton
dynamics \cite{Schulman}, as well as in the spreading of forest fires
\cite{Mendez}.

\begin{figure}[H!]
 \includegraphics[width=0.9\textwidth,angle=0,clip=,]{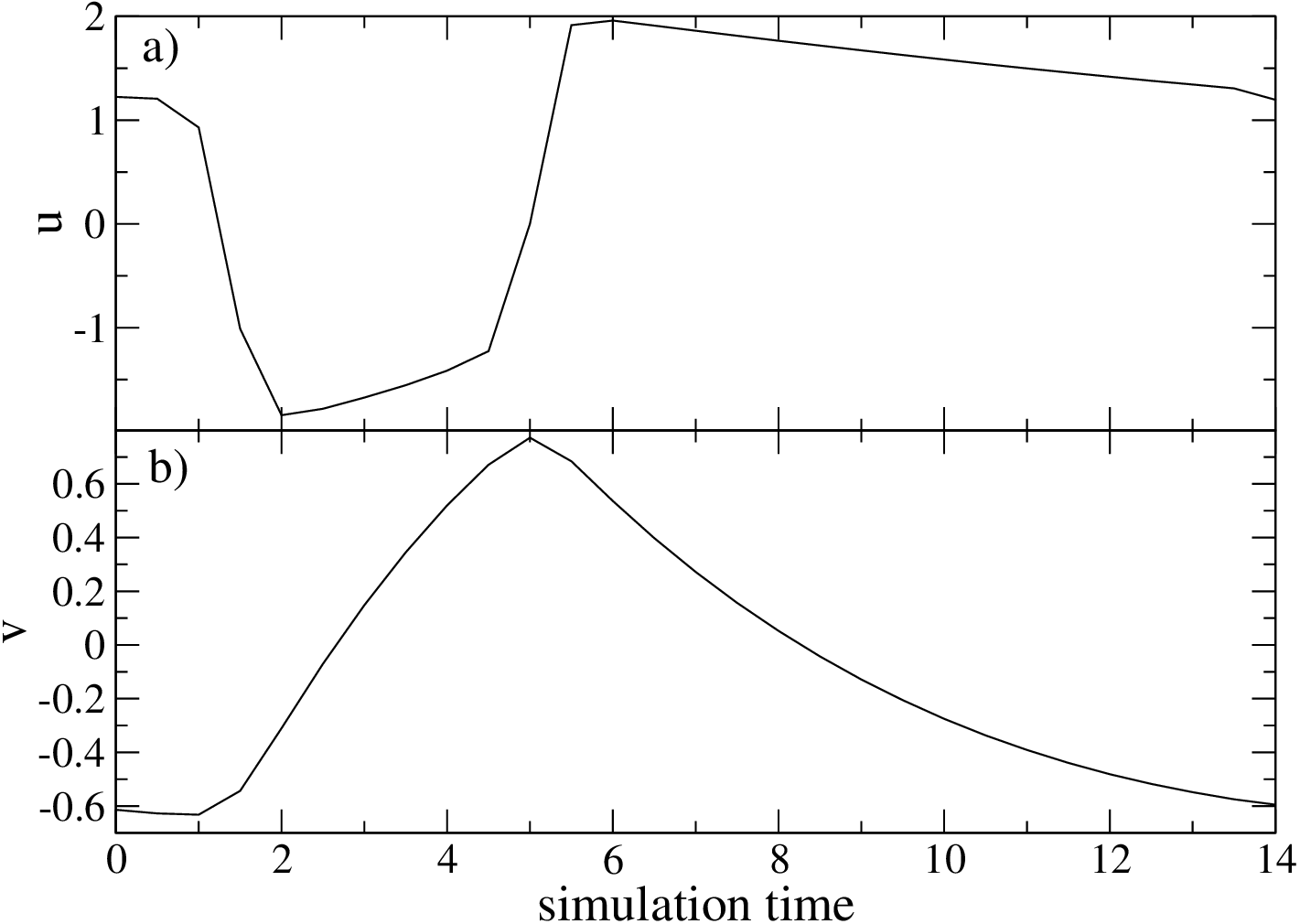}
 \caption{Time evolution of a) $u$ and b) $v$ calculated with the FHN
   equations and the parameters representing a healthy tissue.}
\label{fig:fig1}
\end{figure}

Here we will use Eqs.~(\ref{eq:FHN}) in their original context as a
model to investigate the spatio-temporal evolution of electric
excitations in the heart. In this approach the variable $u$ is roughly
associated with the membrane potential and the variable $v$ with the
ion currents through the cell membrane. The resting state is given by
the pair of values $u=u_0=1.2$ and $v=v_0=-0.6$. The diffusion
coefficient $D$ describes the coupling between the cells, and $z$ is
an applied external current (stimulus). The influence of the
parameters $a$, $b$ and $c$ can be inferred by numerical solutions of
Eqs.~(\ref{eq:FHN}) without the diffusive term. The parameter values
have to be limited to some range in order to generate excitability,
and their detailed effect on the pulses is complicated due to mutual
interdependencies originating from the nonlinearity in
Eq.~(\ref{eq:FHN}). Roughly speaking, $a$ affects the length of the
refractory period, $b$ influences the stability of the resting state,
and $c$ controls the excitability and strength of the cells' response
to a stimulus. To capture the propagation and form of a typical action
potential, the following set of parameters can be used: $D=D_0=0.1$,
$a=a_0=0.7$, $b=b_0=0.6$, and $c=c_0=5.5$. These values will be
associated with a ``healthy tissue'' in the following.
Figure~\ref{fig:fig1} shows the time development of $u$ and $v$ during
an excitation after a stimulus with these parameters. Note that the
variable $-u$ mirrors the form of a pulse in an usual representation
of an ECG recording.

Ectopic foci and spiral waves are thought to be caused and influenced
by physiologically modified regions of the tissue, which in the
modeling correspond to spatial variations of the parameters. To
simplify the analysis, we fix $a=a_0$ and $D=D_0$, and consider
variations of the parameters $b$ and $c$ according to \begin{eqnarray}
\label{eq:obstacle b} b(x,y) &=& b_0 - \Delta
b\exp(-\sqrt{(x-x_0)^2+(y-y_0)^2}/\xi_b)\,,\\ \label{eq:obstacle c}
c(x,y) &=& c_0 - \Delta c\exp(-\sqrt{(x-x_0)^2+(y-y_0)^2}/\xi_c)\,,
\end{eqnarray} where the amplitudes $\Delta b$, $\Delta c$
characterize the strength, and the correlation lengths $\xi_b$ and
$\xi_c$ characterize the spatial range of modification.

The calculations are carried out on a two-dimensional simulation area
of size $20\times20$, which represents an isolated section of atrial
heart tissue, as it is used often in experiments
\cite{Ikeda_2,Iravanian,Ikeda_3}. The boundary conditions of the
simulation area are of von Neumann type, i.\ e.\ $\partial u/\partial
n=0$, where $\partial/\partial n$ denotes the normal derivative.

To solve the two nonlinear coupled partial differential equations
(\ref{eq:FHN}) we use the finite element method (FEM) with a
triangulation consisting of 4225 nodes and 8192 triangles, and a
constant integration time step $\Delta t = 0.01$. A simulation time of
$1$ corresponds to a time of roughly $5$ to $5.5$ ms. The nonlinearity
$u^3(\vec{x},t)$ in Eq.~(\ref{eq:FHN}) is treated as an inhomogeneity,
which means that for $u(\vec{x},t_{i})$ the value $u(\vec{x},t_{i-1})$
of the preceding time step is used.

\section{Generating mechanisms}
\label{ch:mechanisms}
\subsection{Ectopic activity}\label{subsec:ea}

Ectopic foci are regions in the atria, which generate activation waves
emanating from self-excitatory hyperactive cells. In these cells the
transmembrane potential raises without external stimulation until the
threshold value is reached and an action potential results. In optical
mapping studies and spatially resolved ECG recordings, ectopic foci
are often localized in the regions of the pulmonary veins
\cite{Sanders, Sahadevan}.

\begin{figure}[H!]
\begin{minipage}{0.47\textwidth}
 \includegraphics[width=0.99\textwidth,angle=0,clip=,]{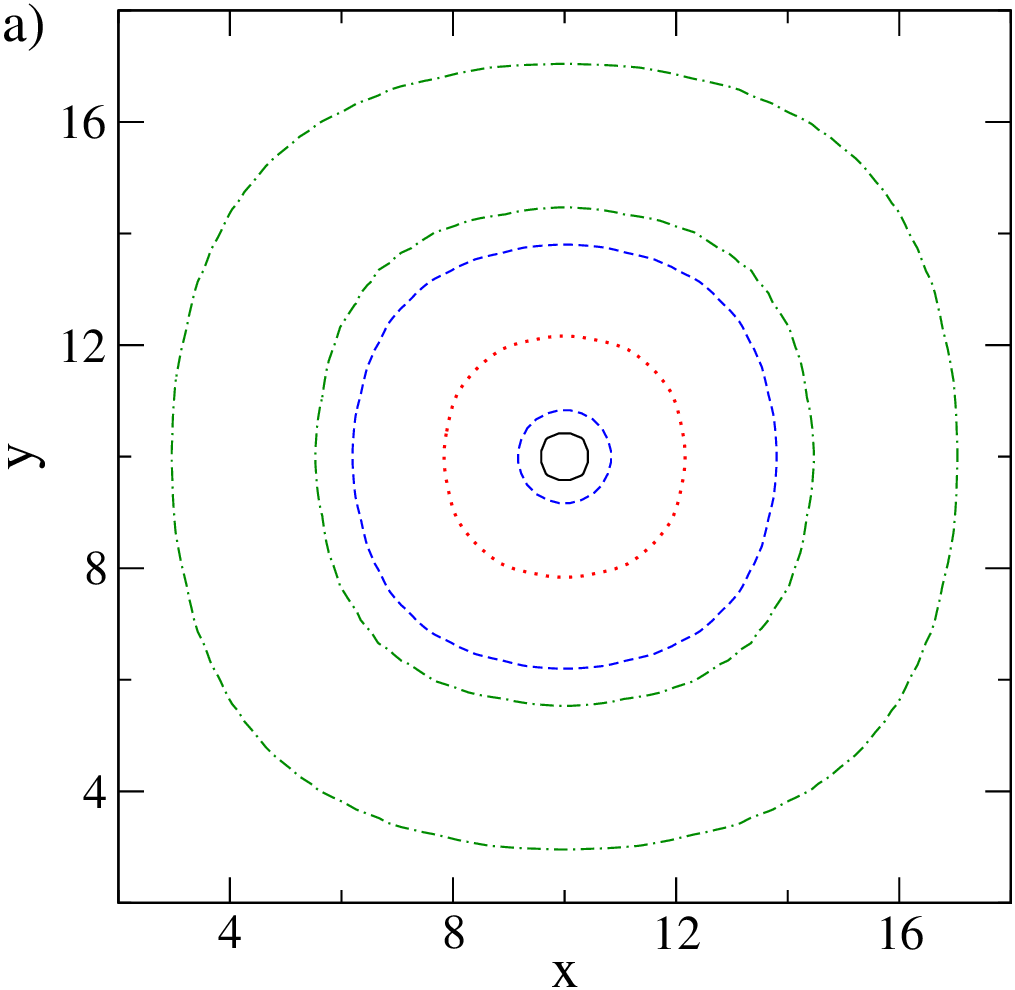}
\end{minipage} 
\begin{minipage}{0.49\textwidth}
\vspace{0.4cm}
\includegraphics[width=0.99\textwidth,angle=0,clip=,]{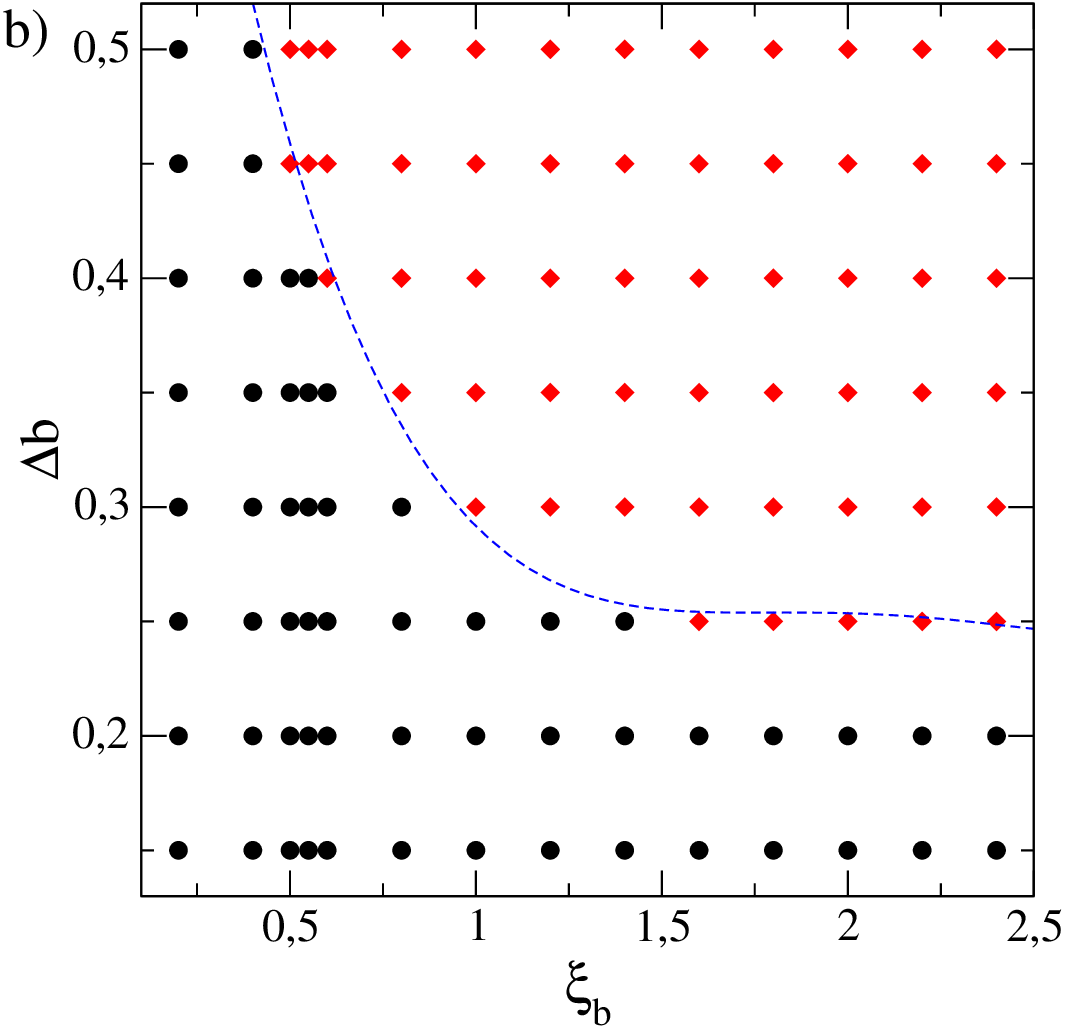}
\end{minipage}
\caption{(color online) a) Activation pattern of the ectopic focus at
  $\Delta b=0.4$ and $\xi_b=0.8$. The lines represent isolines for
  $u=-0.8$ and four times $t=5$ (black solid line), 7 (red dotted
  line), 9 (blue dashed line), and 13 (green dash-dotted line).
  Initially ($t=0$) the system is in the resting state ($u=u_0$,
  $v=v_0$). b) Phase diagram of ectopic activity for modifications
  according to Eq.~(\ref{eq:obstacle b}). Red diamonds and black
  circles refer to the occurrence and absence of ectopic activity,
  respectively. The dotted line is drawn as a guide to the eye and
  marks the transition between the regions of ectopic activity and
  absence of self-excitatory behavior.}
\label{fig:fig2}
\end{figure}

\begin{figure}[H!]
\includegraphics[width=0.7\textwidth,angle=0,clip=,]{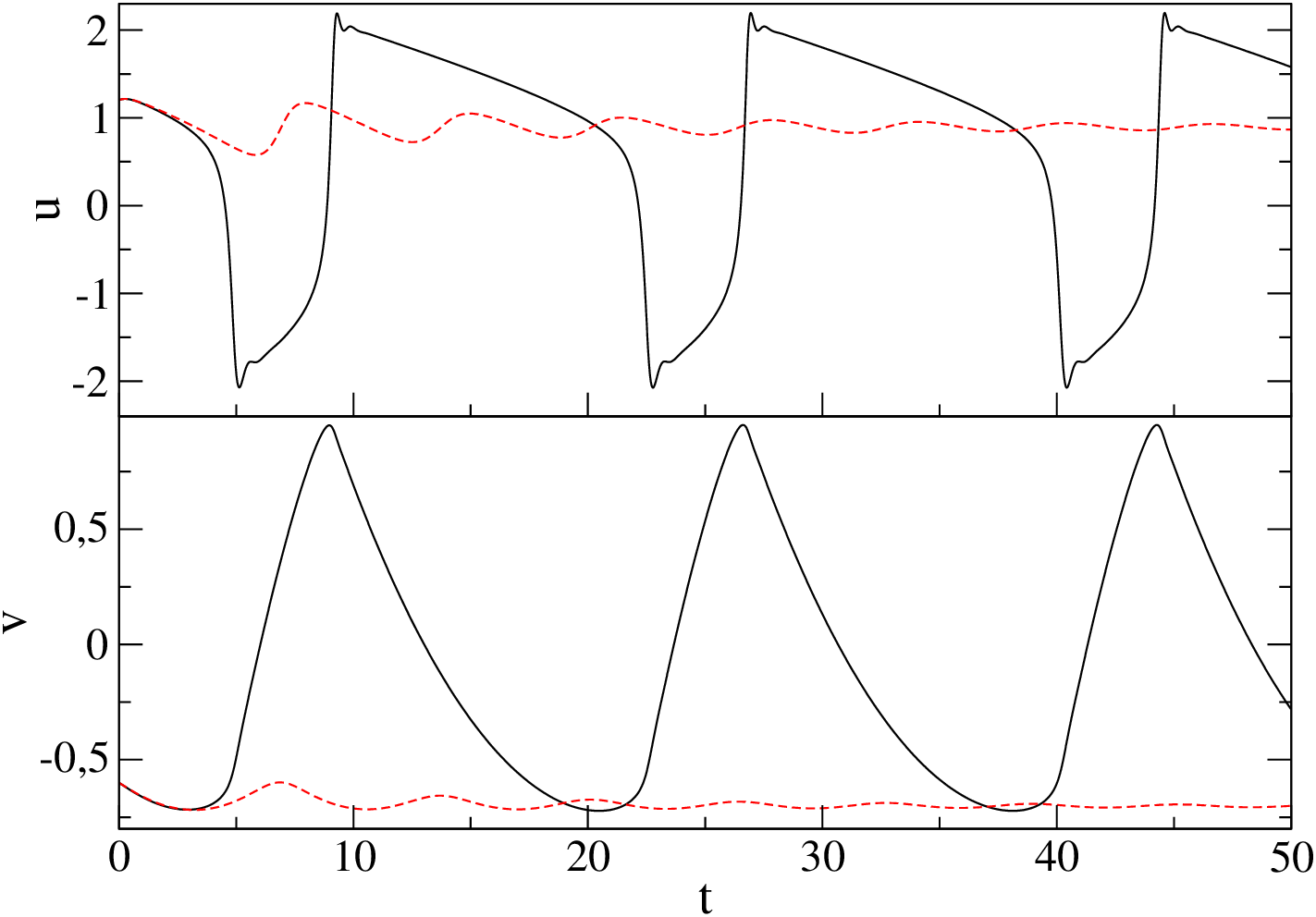}
\caption{(color online) Time evolution of $u$ and $v$ for $\Delta
  b=0.4$, and two values $\xi_b=0.8$ (black solid line) and
  $\xi_b=0.5$ (red dashed line). For $\xi_b=0.8$, $u$ decreases below
  the threshold value $u_{\rm th}\simeq 0.6$ and an action potential
  is generated. For $\xi_b=0.5$ the decrease of $u$ is slower and less
  steep due to the stronger diffusive current compared to $\xi_b=0.8$.
  As a consequence the response of $u$ is more susceptible to the
  initial decrease of $v$, which is not distinguishable for
  $\xi_b=0.5$ and $\xi_b=0.8$. Thus, $u$ does not reach the threshold
  value and relaxes with small oscillations to a value $u\simeq0.9$.}
\label{fig:fig3} 
\end{figure}

To model a tissue with physiologically modified properties that result
in ectopic activity, we fix $c=c_0$ ($\Delta c=0$) and vary the
resting state stability $b$ around the center of the simulation area
($x_0=y_0=10$) according to Eq.~(\ref{eq:obstacle b}). Initially the
system is in the excitable resting state ($u=u_0$ and $v=v_0$).
Figure~\ref{fig:fig2}a shows the resulting activation pattern for
$\Delta b=0.4$ and $\xi_b=0.8$. The modified tissue is self-excitatory
and acts as a pacemaker for activation waves, which propagate
radially. In the time evolution shown in Fig.~\ref{fig:fig3}, $u$
decreases until the threshold value $u_{\rm th}\simeq0.6$ for
activation is reached and an action potential with a steep fall in $u$
occurs. In response to this activation, the inhibitor variable $v$
increases and pulls $u$ back to a value even larger than its initial
value (overshoot) before $u$ returns to it, and the self-excitatory
process starts anew.

In order to systematically characterize the occurrence of ectopic
activity, we calculate a phase diagram, where in dependence of $\xi_b$
and $\Delta b$ regions of ectopic activity can be distinguished from
that without self-excitatory behavior. The results in
Fig.~\ref{fig:fig2}b show that there exists a minimal $\Delta b_{\rm
  min}\simeq 0.25$, below which no ectopic activity occurs. The
corresponding value $(b_0-\Delta b_{\rm min})\simeq 0.5$ is the
critical value of resting state stability in the FHN
equations~(\ref{eq:FHN}) in the absence of the diffusion term. For
fixed $\Delta b>\Delta b_{\rm min}$ the ectopic activity vanishes,
when $\xi_b$ falls below the dashed transition line in
Fig.~\ref{fig:fig2}b. In this region the diffusive current from the
modified tissue to the surrounding causes the initial decrease of $u$
to become so slow, that the counter-regulation by $v$ eventually
hinders $u$ to reach its activation threshold, cf.\
Fig.~\ref{fig:fig3}. Only small oscillation of $u$ around a
reduced resting state value can be seen in Fig.~\ref{fig:fig3},
which become weaker with growing time.

\begin{figure}[H!]
\includegraphics[width=0.7\textwidth,angle=0,clip=,]{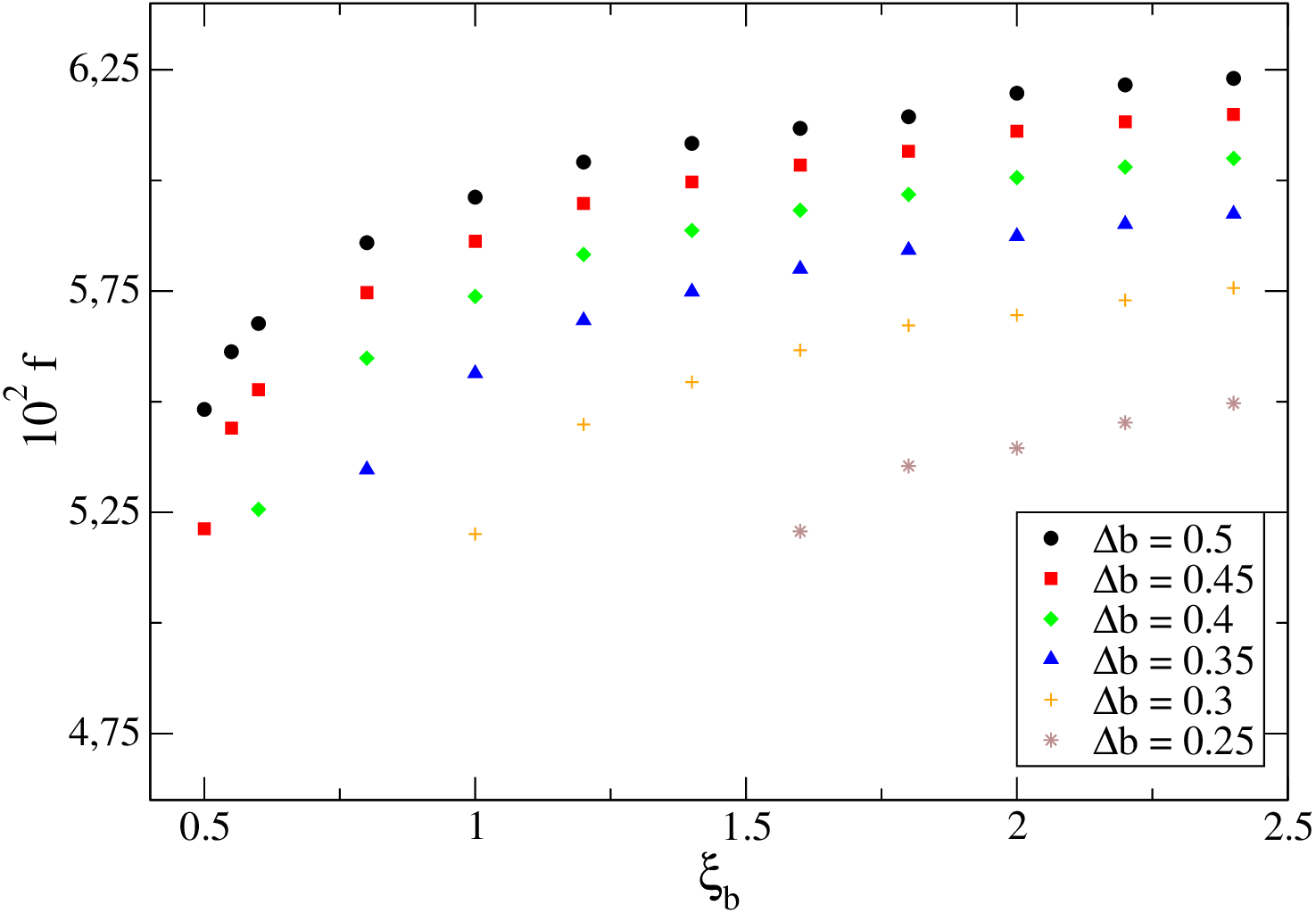}
\caption{(color online) Frequency of ectopic activity in dependence of
  the size $\xi_b$ of the modified tissue for various modification
  strength $\Delta b$ of the resting state stability. A frequency of
  $0.053$ in the simulation corresponds to a frequency of roughly $10$
  Hz.}
\label{fig:fig4}
\end{figure}

The temporal-spatial pattern of the activation in the phase of ectopic
activity is characterized by the frequency of the ectopic focus. We
calculate this frequency as the inverse mean time interval between
consecutive action potentials. As shown in Fig.~\ref{fig:fig4}, the
frequency becomes larger with increasing $\Delta b$ (at fixed $\xi_b$)
and $\xi_b$ (at fixed $\Delta b$), and it tends to saturate for large
$\xi_b$. With increasing $\xi_b$, the diffusive current of the inner
cells of the modified tissue decreases and thus a larger frequency is
obtained. In the saturation limit the frequency is nearly the same as
in the absence of diffusion and thus is mainly determined by the
refractory period.

\subsection{Spiral waves}
\label{ch:spiral waves} 

In this section we study the influence of physiologically modified
regions, called ``obstacles'' henceforth, on spiral wave behavior. It
was observed that spiral waves in the atria can be generated by a
perturbation of the propagation of planar excitation waves by anatomic
obstacles as, for example, the pulmonary veins, the venae cavae, the
pectinate muscle bundles or some localized region of modified tissue
\cite{Wu, Lim, Ikeda_2,Pertsov}. These regions are considered as not
fully excitable and are thus modeled as regions with a reduced
parameter $c$ according to Eq.~(\ref{eq:obstacle c}).

In an experiment by Ikeda and coworkers \cite{Ikeda_2} a nearly
rectangular area of atrial tissue was placed on an electrode plaque in
a tissue bath. Holes with different diameters were created and a
reentrant wave was initiated by cross-field stimulation. The resulting
behavior of the wavefront was, amongst others, classified according to
whether the spiral is anchored by the obstacle, and by the
relationship between hole size and cycle length of the reentry. It was
observed that for large obstacle sizes (6, 8 and 10 mm), the reentrant
wave attaches to the obstacle, leading to a linear increase of the
cycle length with the hole diameter. For small obstacle diameters
below about 4~mm, by contrast, meandering spirals with a tip getting
variably closer to or further away from the hole were found. In this
case the cycle length becomes independent of the hole diameter.
Similar results were observed by Lim and coworkers \cite{Lim}. They
analyzed the behavior of spiral waves near holes with diameters
ranging from $0.6$ to $2.6$~mm and obtained a higher attachment rate
for larger obstacle diameters as well as a positive linear correlation
of the reentry conduction velocity and wave length with the obstacle
diameter in the case of attached spirals For smaller obstacle
diameters the spiral waves were found to attach to and detach from the
obstacle.

The missing anchoring for small hole sizes was explained in
\cite{Ikeda_2} by invoking a ``source-sink relationship''. The
``source'' is the activation wavefront and provides a diffusive
current to the surrounding tissue in the resting state, which
constitutes the ``sink''. The sink becomes larger for smaller
obstacles, where more cells become depolarized by the activation
wavefront. If the source-to-sink ratio is decreased below a certain
critical value, the wavefront detaches from the obstacle.

\begin{figure}[H!]
\begin{minipage}{0.48\textwidth}
\includegraphics[width=0.99\textwidth,angle=0,clip=,]{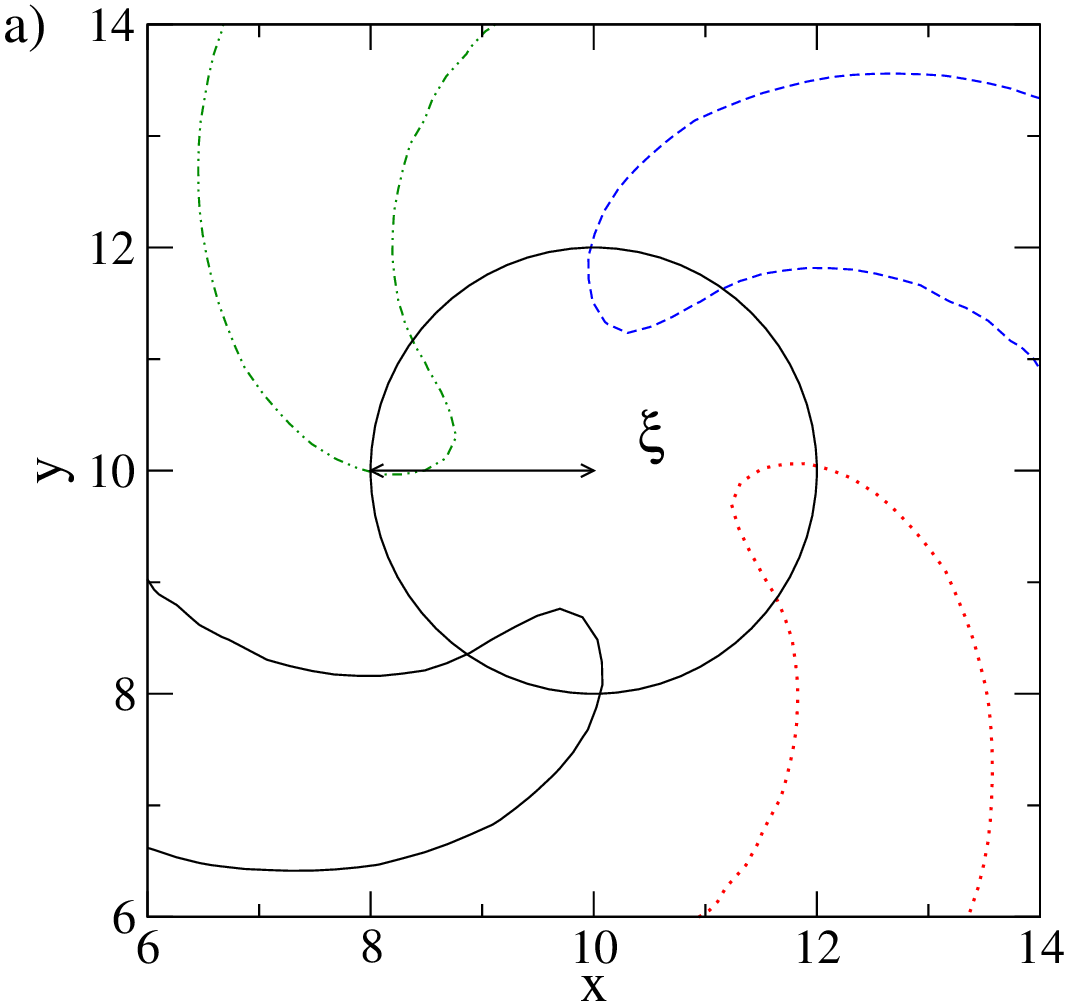}
\end{minipage}
\begin{minipage}{0.48\textwidth}
 \includegraphics[width=0.99\textwidth,angle=0,clip=,]{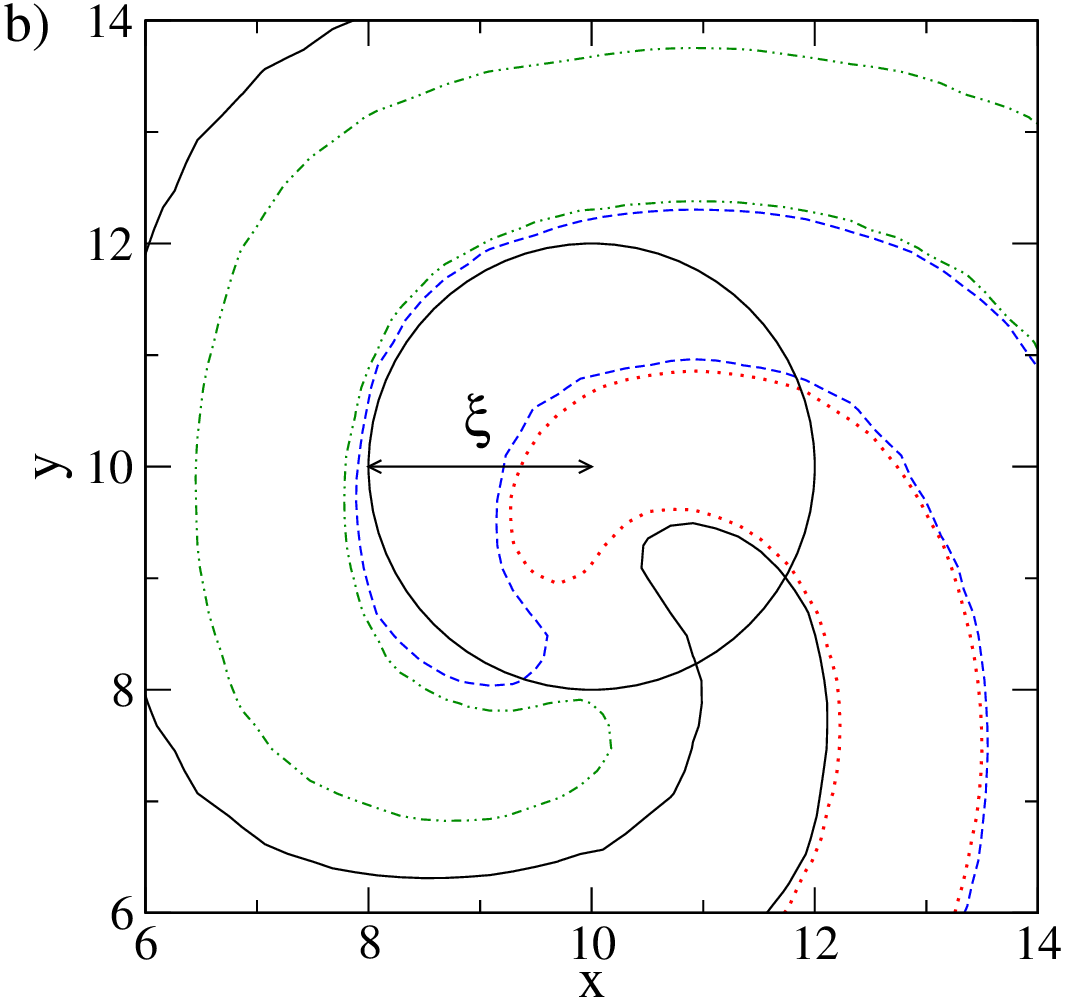}
\end{minipage}
\caption{a) Isolines for $u = -0.8$ at four different times $t = 82$
  (black solid line), $87$ (red dotted line), $92$ (blue dashed line)
  and $t = 97$ (dash-dotted line) for an obstacle with $\xi_c=2$ and
  $\Delta c=4.5$ (marked by the black circle). The spiral wave is
  pinned and rotates around the obstacle (anatomical reentry). b)
  Isolines for $u = -0.8$ at four different times $t = 82$ (black
  solid line), $84.5$ (red dotted line), $87$ (blue dashed line) and
  $t = 89.5$ (dash-dotted line) for an obstacle with $\xi_c=2$ and
  $\Delta c=1.5$. The spiral wave rotates around a moving center not
  corresponding to but influenced by the obstacle and is not attached
  to the obstacle (functional reentry).}
\label{fig:fig5}
\end{figure}

To elucidate these experimental findings, we perform numerical
calculations for a geometry corresponding to the experiments with the
following initial state and parameters settings: the modified region,
is, as in the previous Sec.~\ref{subsec:ea}, placed in the center of
the simulation area at $x_0=y_0=10$. Initially a ``planar'' (linear)
wave is generated by inducing a current $z$ in the stripe $9.5 \leq
x\leq10$, $0\leq y \leq10$, and by setting the area $0\le x\le9.5$,
$0\le y\le10$ into a refractory state with $u=1.6$ and $v=0$, while
the rest of the simulation area is in the resting state ($u=u_0$,
$v=v_0$). This initial state resembles the activation pattern directly
after application of a cross-field or paired-pulse stimulation (two
rectangular pulses). At the ``upper part'' of the initial planar
wavefront ($9.5\le x\le 10$, $y=10$), diffusive currents flow
``radially'' in all forward directions ($y>10$), while at the ``right
boundary'' ($x=10$, $0\le y\le 9.5$) the diffusive currents can flow
only in positive $x$ direction (due to the refractory state in the
area $0\le x\le 9.5$, $0\le y\le 10$). This higher loss by diffusion
leads to a smaller propagation speed of the initial wavefront at its
upper boundary compared to its right boundary. As a consequence, the
wavefront becomes curved, and a reentrant spiral wave develops for all
reductions $\Delta c$ in excitability and obstacle sizes $\xi_c$, in
accordance with the experimental observations. Figure~\ref{fig:fig5}
shows activation patterns for $\xi_c=2$ and a) $\Delta c=4.5$, and b)
$\Delta c=1.5$. The stronger reduction of excitability in
Fig.~\ref{fig:fig5}a leads to an anchoring of the spiral wave, while
in Fig.~\ref{fig:fig5}b the spiral is meandering.

\begin{figure}[H!]
\includegraphics[width=0.7\textwidth,angle=0,clip=,]{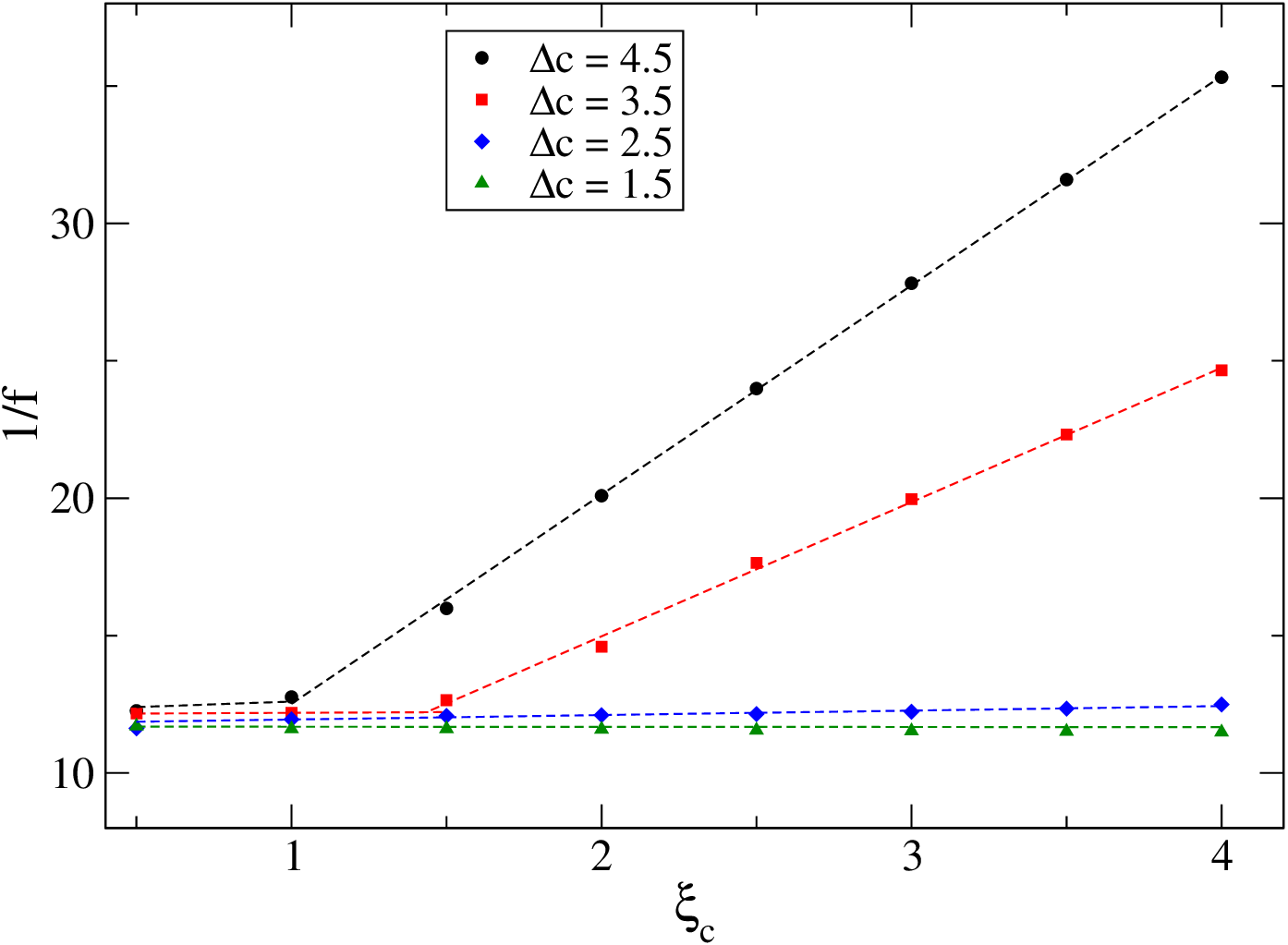}
\caption{(color online) Mean cycle length $1/f$ as a function of
  obstacle size $\xi_c$ for four different strength of the
  modification of excitability $\Delta c$. A cycle length of $15$
  corresponds to a frequency of roughly $12$ Hz. For the two smaller
  values $\Delta c=1.5$ and $\Delta c=2.5$ the cycle length is
  independent of $\xi_c$ (functional reentry). For the two larger
  values $\Delta c=3.5$ and $\Delta c=4.5$ a transition from
  functional to anatomical reentry occurs, when $\xi_c$ exceeds a
  threshold value $\xi_c^\star$ that increases with decreasing $\Delta
  c$. The dashed lines are fits to the data for mean cycle lengths
  independent of $\xi_c$ (functional reentry) and mean cycle lengths
  proportional to $\xi_c$ (anatomical reentry).}
\label{fig:fig6}
\end{figure}

To analyze the parameter regimes of the occurrence of anchored or
meandering spiral waves, we perform a frequency analysis for different
values of $\Delta c$ and $\xi_c$. Therefore, we determine the peak
positions in the time series of $u$ at $8$ positions far away from the
center of the spiral and calculate the peak-to-peak intervals. The
frequency of one point is one over the mean of the peak-to-peak
intervals and the mean cycle length is one over the average of all
these local frequencies. The results in Fig.~\ref{fig:fig6} show that,
as in the experiments, attached spiral waves occur for large $\Delta
c\gtrsim3$ and for sufficiently large $\xi_c>\xi_c^\star$, where
$\xi_c^\star$ decreases with increasing $\Delta c$. For these anchored
spirals, the frequency is proportional to $f=\eta/2\pi\xi_c$, where
$\eta\simeq 0.82$ is the conduction velocity in the FHN model.
Accordingly, $1/f$ increases linearly with $\xi_c$ for
$\xi_c>\xi_c^\star$ in Fig.~\ref{fig:fig6}. For small $\Delta
c\lesssim3$, only meandering spirals are observed. The transition from
large to small $\Delta c$ reflects the transition from anatomical to
functional reentry \cite{Boermsa}, as it has been reported in medical
studies \cite{Lim}. The fact that for small $\xi_c$ always meandering
spirals occur, can be interpreted by the small source-sink ratio
\cite{Ikeda_2}. The same mechanism can also lead to meandering spirals
for large $\xi_c$ if $\Delta c$ becomes small. Note, that nevertheless
the spiral wave is not anchored to the obstacle, its movement is still
influenced by the obstacle.

\section{Induced fibrillatory states in the right atrium}
\label{ch:induction}

In previous studies on interactions of paced waves with
self-excitatory waves, the influence of the pacing on a spiral wave
was studied \cite{Osipov,Davidenko} with the aim to suggest a possible
therapy to suppress fibrillation or tachycardia. The pacing was
applied to the region, where the spiral wave was located. It was found
that the pacing leads to an annihilation of the reentrant activity or
to a shift of the spiral core \cite{Davidenko,Agladze,Fu,Gottwald}.

\begin{figure}[H!]
 \includegraphics[width=0.99\textwidth,angle=0,clip=,]{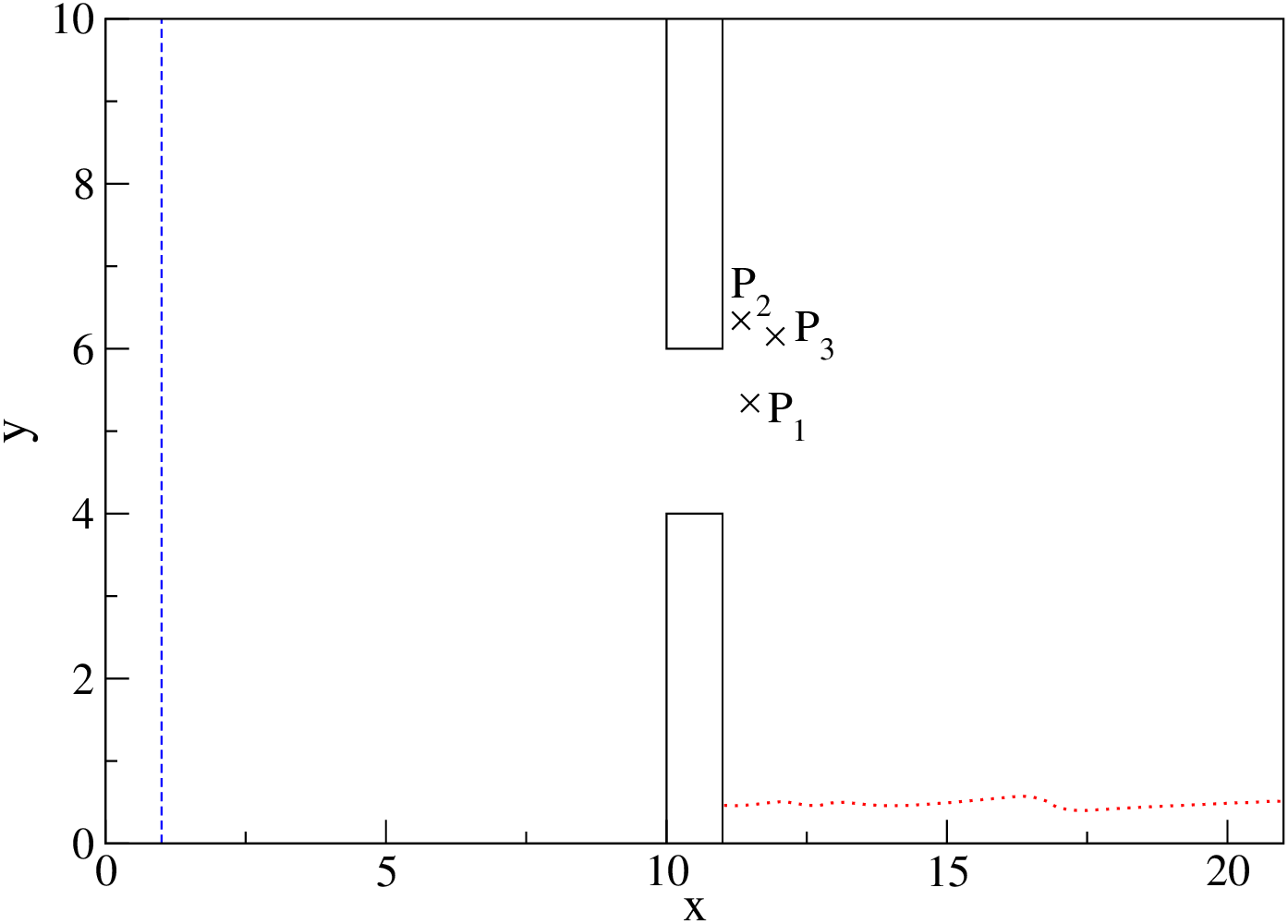}
 \caption{Illustration of the simulation area. The red dotted line
   describes a regular paced wavefront of the sinus node. The blue
   dashed line is a wavefront of the perturbing pacemaker. P$_1$,
   P$_2$ and P$_3$ mark the "observation points", for which the time
   evolution of $u$ is shown in Figure~\ref{fig:fig9}.}
\label{fig:fig7}
\end{figure}

Here we investigate the perturbation of regular paced waves from a
source representing the sinus node by waves emanating from an
additional pacemaker located in a distant region. Electrocardiogram
recordings and their frequency analyses show that regular excitation
patterns are often observed in the left atrium, where additional
pacemakers like spiral waves or ectopic foci are located
\cite{Sanders, Mandapati}, and that at the same time irregular,
fibrillatory-like states in the right atrium occur \cite{Sahadevan}.
This led to the conjecture that fibrillatory states can be induced in
the right atrium by self-excitatory pacemakers in the left atrium. In
this connection it is important to better understand how a
fibrillatory state can occur, if regular paced waves, as generated by
the sinus node, are disturbed by additional pacemaker waves. To this
end, we consider the waves to be located in spatially separated
regions that are connected by a small region. To be specific, we
choose a simulation area of size $21\times10$, which is divided into
three regions (see Fig.~\ref{fig:fig7}). The rectangular area L
with $0\leq x \leq 10$, $0\leq y\leq 10$ representing the left atrium,
the rectangular area R with $11\leq x \leq 21$, $0\leq y\leq 10$
representing the right atrium, and the small bridge B with $10<x<11$,
$4<y<6$ representing the connection between the atria. The grid used
in the finite element calculations consists of in total 8871 nodes and
17350 triangles.

We focus on situations where the pacemaker in the region L is located
far outside the left part of the simulation area, so that the
resulting wavefronts become ``planar'' (linear). In the simulation
they are generated by application of a stimulating current $z=-1$ with
duration $t_z =1 = 100 \Delta t$ and a period $1/f_{\rm pert}$ in the
region $x\leq 0.5$ and $0 \leq y\leq 10$. The activation waves
representing the pacemaker in region R are generated by the
application of a current $z=-1$ with duration $t_z =1$ and period
$1/f_{\rm pace}$ in the region $11\leq x\leq 21$ and $y\leq 0.5$.

The irregularity of the resulting patterns in region R is quantified
by calculating the Shannon entropy of the distribution of local
activation frequencies for every grid point in R. To this end we
divide the frequency range into $N_{\rm b}$ bins of size
$\Delta=$min$\left\lbrace \exp\left[0.626+0.4\ln(N_{g}-1)
  \right]^{-1},0.01 \right\rbrace $ \cite{Tass,Rosenblum} and
calculate the probabilities
\begin{equation}
p_l = \frac{n(f_l\leq f \leq f_l+\Delta)}{N_{\rm g}}\, ,
\label{eq:probability} 
\end{equation} 
of finding frequency $f$ in bin $l$, where $N_{\rm g}$ is the total
number of grid points. The normalized entropy then is given by 
\begin{equation}
\label{eq:entropiemass} 
s =\frac{S}{S_{max}}=-\frac{\sum_{l=1}^{N_{\rm b}}p_l\ln p_l}{\ln N_{\rm b}}\;\, , 
\end{equation}
For a single frequency ($p_l=\delta_{l,l_0}$), $s=0$, while for a
chaotic activation pattern with a uniform distribution ($p_l=1/N_{\rm b}$),
$s=1$.

For small perturbation frequencies ($f_{\rm pert}\le0.1$) the
influence of the activation wavefronts from the additional pacemaker
onto the sinus node waves is almost negligible. Small deformations of
the linear wavefronts are observed, but the measured frequencies are
close to the pacing frequency, and the overall spatiotemporal pattern
in R is regular.
\begin{figure}[H!]
\begin{minipage}{0.48\textwidth}
 \includegraphics[width=0.99\textwidth,angle=0,clip=,]{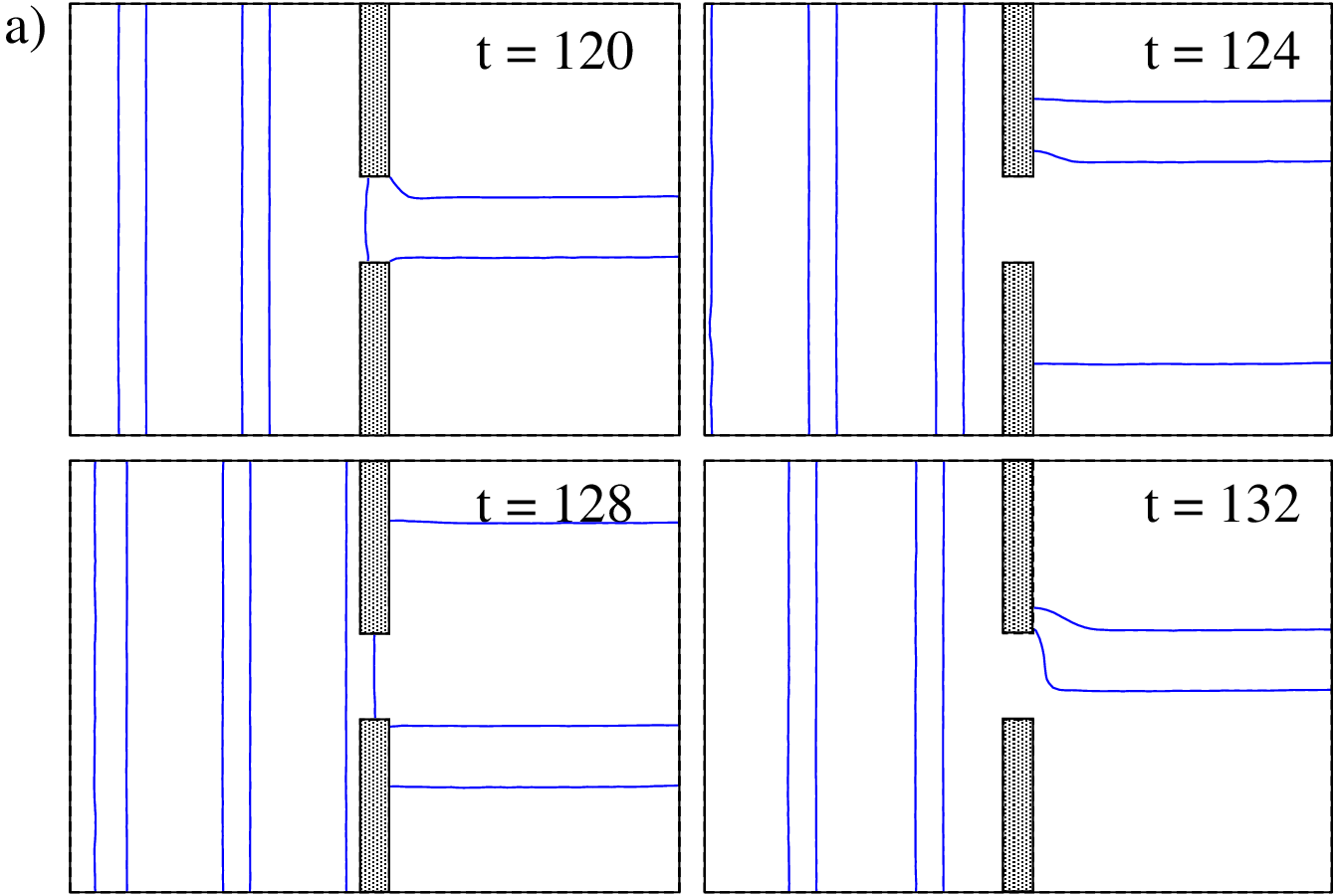}
\end{minipage}
\begin{minipage}{0.48\textwidth}
 \includegraphics[width=0.99\textwidth,angle=0,clip=,]{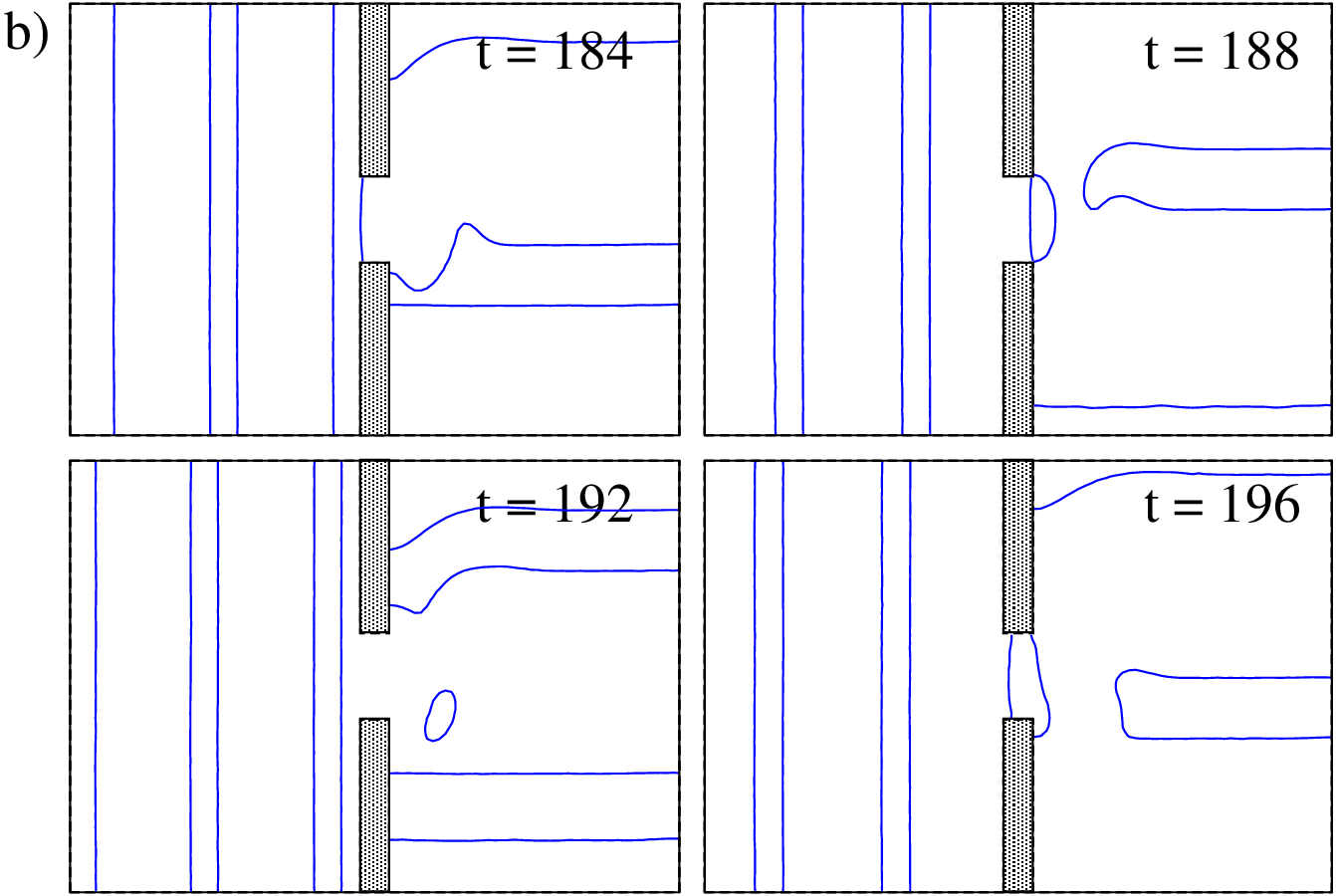}
\end{minipage}
\caption{Time evolution of excitation pattern observed by perturbation
  of a regular pacemaker with frequency $f_{\rm pace}=0.091$ in the
  right part R of the simulation area by an additional pacemaker with
  frequency $f_{\rm pert}= 0.105$ in the left part L through a bridge
  region B. The blue solid lines are isolines for $u=-0.8$. The shaded
  regions represent the boundaries of the bridge between the two parts
  L and R. a) For small times the excitation in R is regular. b) At
  later times breakups of waves in region R occur close to the bridge
  region B, resulting in irregular excitation patterns.}
\label{fig:fig8}
\end{figure}

With increasing perturbation frequency the spatiotemporal pattern in
region R becomes more irregular and a breakup of the regularly paced
waves can occur. Figure~\ref{fig:fig8} shows the time evolution of the
excitation patterns for a perturbation frequency $f_{\rm pert} =
0.105$ and a pacing frequency $f_{\rm pace}=0.091$. For small times,
the waves are only slightly perturbed and the pattern remains regular,
as can be seen from the four consecutive snapshots in
Fig.~\ref{fig:fig8}a. At a later time, however, the perturbation by
the waves from region L results in a breakup of the waves close to the
bridge B in region R, as can be seen from the four consecutive
snapshots in Fig.~\ref{fig:fig8}b. The onset of this breakup was found
to occur at a time $t\simeq160$.

\begin{figure}[H!]
\includegraphics[width=0.7\textwidth,angle=0,clip=,]{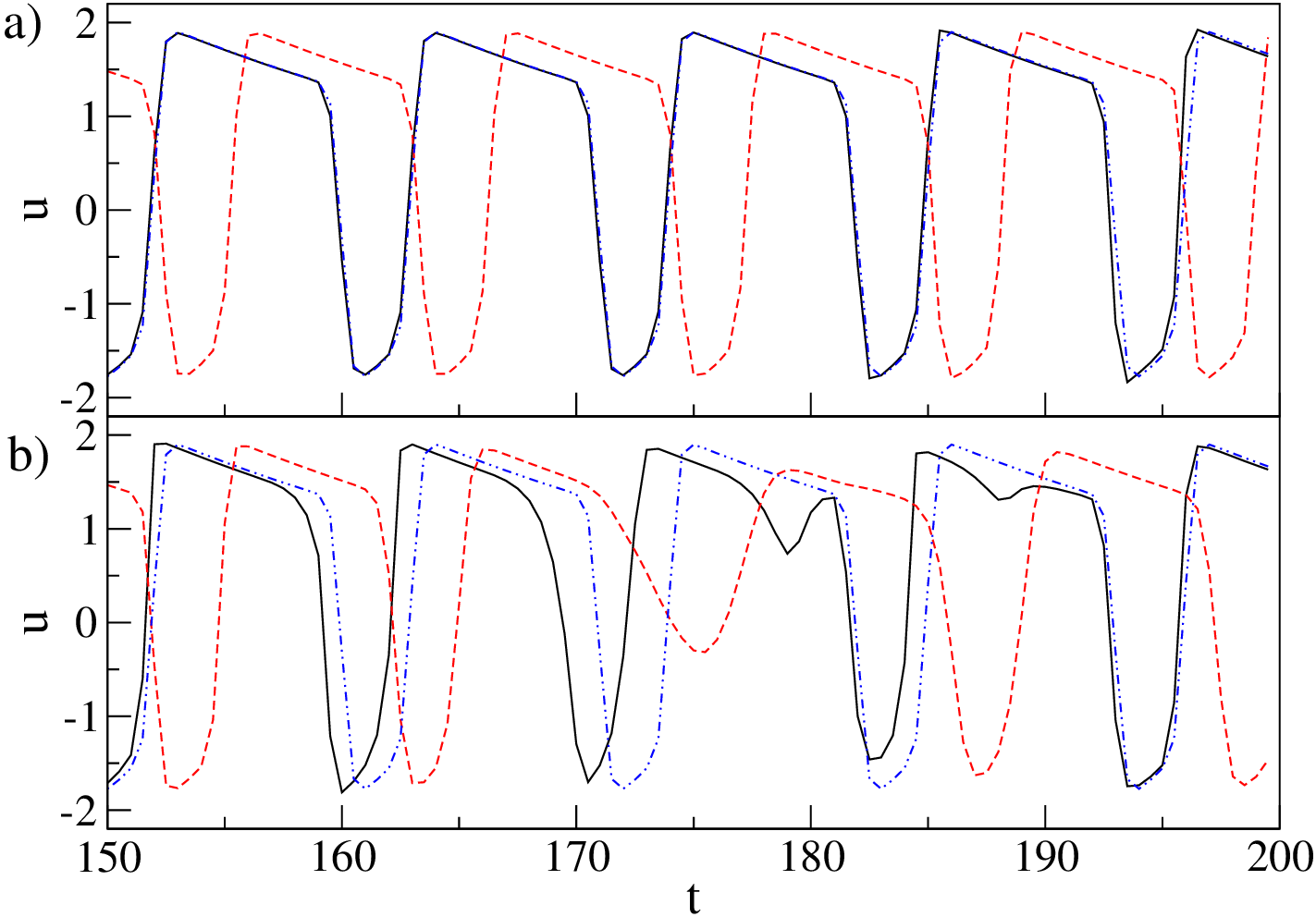}
\caption{(color online) Time evolution of the activation variable $u$
  at three different points of the simulation area: $P_1=(11.49,
  5.34)$ (black solid line), $P_2=(11.33, 6.34)$ (red dashed line) and
  $P_3=(11.94, 6.15)$ (blue dash-dotted line) for a pacing frequency
  $f_{\rm pace}=0.091$ and two different perturbation frequencies a)
  $f_{\rm pert} = 0.1$, and b) $f_{\rm pert} = 0.105$. We have checked
  that this behaviour remains qualitatively the same even at a three
  times longer simulation time which suggests that this time evolution
  corresponds to a stationary state.}
\label{fig:fig9}
\end{figure}

In order to investigate how the spatial irregularity is reflected in
the time evolution, we consider three different points $P_1= (11.49,
5.34), P_2 = (11.33, 6.34)$ and $P_3 = (11.94, 6.15)$ in region R. The
time evolution of $u$ for these three points is shown for two
different perturbation frequencies $f^{(1)}_{\rm pert} = 0.1$ and
$f^{(2)}_{\rm pert} = 0.105$ in Fig.~\ref{fig:fig9} ($P_1$: black
solid curve, $P_2$: red dashed curve, $P_3$: blue dash-dotted curve).
For the lower frequency $f^{(1)}_{\rm pert}$ the evolution at all
three points is regular, see Fig.~\ref{fig:fig9}a. For the higher
frequency $f^{(2)}_{\rm pert}$, by contrast, the break ups of the
waves seen in Fig.~\ref{fig:fig8}b yield unsuccessful activations
during refractory periods, as can be seen, for example, at time
$t\simeq175$ in point P$_2$ (red dashed curve) and at time $t=179$ in
point P$_1$ (black solid curve). These unsuccessful activations are
caused by a rapid pacing of the region by the curled wave. Another
feature is that the shape of the action potential varies. This can be
seen, for example, at point P$_1$ (black solid line) when comparing
the pulses at $t\simeq160$ and $t\simeq194$. It is important to note
that these irregularities are hardly observed at point P$_3$ (blue
dash-dotted curve), showing that they exhibit a spatial heterogeneity.

\begin{figure}[H!]
\includegraphics[width=0.7\textwidth,angle=0,clip=,]{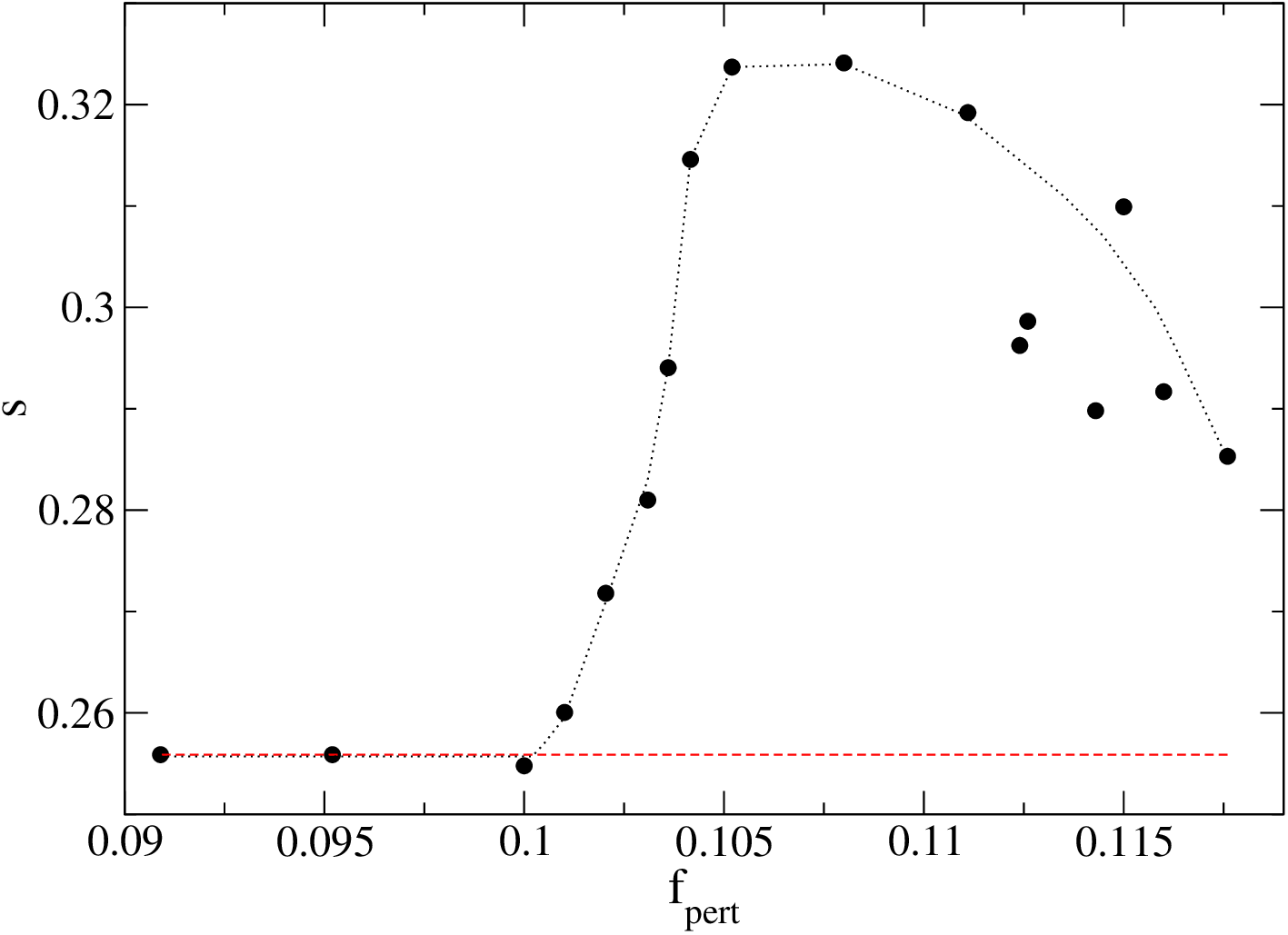}
\caption{(color online) Normalized entropy of the local frequency
  distribution in region R as a function of the frequency $f_{\rm
    pert}$ of the perturbing waves in region L. The pacing frequency
  in region R is $f_\mathrm{pace}=0.091$. The dotted (black) line
  through the data points is a guide to the eye and the dashed line
  marks the value of $s$ for the regularly paced system without
  perturbation by an additional pacemaker.}
\label{fig:fig10} 
\end{figure}

The total irregularity in region R quantified by the normalized
Shannon entropy $s$ of the local frequency distribution is shown in
Fig.~\ref{fig:fig10} as a function of the frequency $f_\mathrm{pert}$
of the perturbing waves from region L. For small perturbation
frequencies $f_\mathrm{pert}\lesssim0.1$ the entropy $s$ equals the
unperturbed case, while for $f_\mathrm{pert}\gtrsim0.1$, $s$ sharply
increases until it reaches a maximum at $f_\mathrm{pert}\simeq0.1075$.
For higher $f_\mathrm{pert}$ a return to more regular activation
pattern is found, indicating that the disturbance is most pronounced
if $f_\mathrm{pert}$ is close to $f_\mathrm{pace}$.

To conclude, the disturbance of the wavefronts in region R by waves
emanating from an additional source in region L and propagating
through the bridge region B can lead to irregular, fibrillatory-like
activation patterns in region R. On the other hand, the waves in
region L are almost unaffected by the waves in R with primary
wavefront orthogonal to the cross section of the bridge. The
irregularities in region R are most pronounced at a certain
perturbation frequency $f_\mathrm{pert}$. How this value is influenced
by the geometry of the bridge and the parameters characterizing the
cell properties remains further investigation.

\section{Summary}
\label{ch:summary} 

The influence of physiologically modified regions on the generation
and properties of spatio-temporal activation patterns was investigated
on the basis of the FitzHugh-Nagumo equations with von Neumann
boundary conditions, in particular the occurrence of ectopic foci and
spiral waves under spatial inhomogeneities of the parameters
characterizing the cell properties. It was shown that the reduction
$\Delta b$ of the resting state stability in circular regions of the
tissue can lead to ectopic activity. A minimal size of hyperactive
tissue is necessary for ectopic activity to occur, as well as a
minimal strength of the reduction of resting state stability with
respect to the ``healthy'' reference value. With increasing size
$\xi_b$ of the hyperactive tissue, the frequency of the ectopic focus
first increases and eventually saturates. The saturation frequency
depends on the strength of the modification $\Delta b$.

For spiral wave patterns it was found that an anchoring of the wave to
the obstacle can occur. To uncover this mechanism, an obstacle was
modeled as a patch of modified tissue with reduced excitability by a
reduction of the parameter $c$ in the FHN equations. The obstacle was
placed in the middle of a two-dimensional square simulation area and a
planar excitation wave was generated aside of the obstacle in front of
a refractory region, which represents an activation pattern observed
after cross-field stimulation in experiments. As in the experiments,
reentrant waves are observed. These exhibit either functional or
anatomical reentry in dependence of the obstacle size and reduction
strength $\Delta c$ of excitability. An analysis of the spiral wave
frequency in dependence of the obstacle size yields results in
accordance with the experimental observations.

Finally we studied the question, if and how fibrillatory-like states
can arise in the right atrium due to the presence of self-excitatory
spiral waves or ectopic foci in the left atrium. To this end the
simulation area was separated into two rectangular regions L and R
connected by a small bridge B. Planar excitation waves were generated
with different frequencies in the left region L to model a pacemaker
far outside the left part of the simulation area. Planar excitation
waves resembling stimulation by the sinus node were generated by
periodic application of a stimulating current at one boundary in the
right part R of the simulation area. For small perturbation
frequencies in L, the disturbance of the waves in R turned out to be
small. For higher perturbation frequencies , the waves in R become
significantly disturbed and the spatio-temporal activation pattern
eventually becomes irregular. The time evolution of the activation
variable $u$, representing the electric potential in the FHN
equations, shows features in close resemblance to the ones found in
intra-atrial electrocardiogram recordings during fibrillation in the
right atrium. The spatial variation of the excitation frequency was
quantified in terms of an entropy, which showed, for a given pacing
frequency, a maximum as a function of the perturbation frequency.
Further investigations will focus on the influence of the geometry of
the bridge and the wavefronts as well as analyse the behavior for
different pacing frequencies. The reliability of the measure of
irregularity $s$ should be analysed and if necessary other methods to
characterise the system behavior should be searched.

\acknowledgments C.\ L.\ thanks the Thuringian government for
financial support.

\end{document}